%Busch and Shimony
%Insolubility of the Quantum Measurement Problem for Unsharp Observables
%Studies in History and Philosophy of Modern Physics

%Contents: (to be found with these headings)

%%%%%%%%%%%%%%%%%%%%%%%%%%%%%%%%%%%%%%%%%%%%%%%%%%%%%%%%%%%%%
%1. input: vanilla.sty

%%%%%%%%%%%%%%%%%%%%%%%%%%%%%%%%%%%%%%%%%%%%%%%%%%%%%%%%%%%%%
%2. input: our macros

%%%%%%%%%%%%%%%%%%%%%%%%%%%%%%%%%%%%%%%%%%%%%%%%%%%%%%%%%%%%%
%3. text

%%%%%%%%%%%%%%%%%%%%%%%%%%%%%%%%%%%%%%%%%%%%%%%%%%%%%%%%%%%%%
%1. input: vanilla.sty

% VANILLA.STY
% COPYRIGHT (C) 1985, 1986 BY MICHAEL SPIVAK
% version date 1/1/86
\catcode`\@=11
\font\tensmc=cmcsc10      %change to CM fonts 3-31-87
%\font\tensmc=amcsc10
\def\smc{\tensmc}

\def\hcorrection#1{\advance\hoffset by #1 }
\def\vcorrection#1{\advance\voffset by #1 }
\def\wlog#1{}
\newif\iftitle@
\outer\def\title{\title@true\vglue 24\p@ plus 12\p@ minus 12\p@
   \bgroup\let\\=\cr\tabskip\centering
   \halign to \hsize\bgroup\tenbf\hfill\ignorespaces##\unskip\hfill\cr}
\def\endtitle{\cr\egroup\egroup\vglue 18\p@ plus 12\p@ minus 6\p@}
\outer\def\author{\iftitle@\vglue -18\p@ plus -12\p@ minus -6\p@\fi\vglue
    12\p@ plus 6\p@ minus 3\p@\bgroup\let\\=\cr\tabskip\centering
    \halign to \hsize\bgroup\smc\hfill\ignorespaces##\unskip\hfill\cr}
\def\endauthor{\cr\egroup\egroup\vglue 18\p@ plus 12\p@ minus 6\p@}
\outer\def\heading{\bigbreak\bgroup\let\\=\cr\tabskip\centering
    \halign to \hsize\bgroup\smc\hfill\ignorespaces##\unskip\hfill\cr}
\def\endheading{\cr\egroup\egroup\nobreak\medskip}
\outer\def\subheading#1{\medbreak\noindent{\tenbf\ignorespaces
      #1\unskip.\enspace}\ignorespaces}
\outer\def\proclaim#1{\medbreak\noindent\smc\ignorespaces
    #1\unskip.\enspace\sl\ignorespaces}
\outer\def\endproclaim{\par\ifdim\lastskip<\medskipamount\removelastskip
  \penalty 55 \fi\medskip\rm}
\outer\def\demo#1{\par\ifdim\lastskip<\smallskipamount\removelastskip
    \smallskip\fi\noindent{\smc\ignorespaces#1\unskip:\enspace}\rm
      \ignorespaces}
\outer\def\enddemo{\par\smallskip}
\newcount\footmarkcount@
\footmarkcount@=1
\def\makefootnote@#1#2{\insert\footins{\interlinepenalty=100
  \splittopskip=\ht\strutbox \splitmaxdepth=\dp\strutbox 
  \floatingpenalty=\@MM
  \leftskip=\z@\rightskip=\z@\spaceskip=\z@\xspaceskip=\z@
  \noindent{#1}\footstrut\rm\ignorespaces #2\strut}}
\def\footnote{\let\@sf=\empty\ifhmode\edef\@sf{\spacefactor
   =\the\spacefactor}\/\fi\futurelet\next\footnote@}
\def\footnote@{\ifx"\next\let\next\footnote@@\else
    \let\next\footnote@@@\fi\next}
\def\footnote@@"#1"#2{#1\@sf\relax\makefootnote@{#1}{#2}}
\def\footnote@@@#1{$^{\number\footmarkcount@}$\makefootnote@
   {$^{\number\footmarkcount@}$}{#1}\global\advance\footmarkcount@ by 1 }

\hyphenation{man-u-script man-u-scripts ap-pen-dix ap-pen-di-ces}
\hyphenation{data-base data-bases}
\ifx\amstexloaded@\relax\catcode`\@=13 
  \endinput\else\let\amstexloaded@=\relax\fi
\newlinechar=`\^^J
\def\eat@#1{}
\def\Space@.{\futurelet\Space@\relax}
\Space@. % 
\newhelp\athelp@
{Only certain combinations beginning with @ make sense to me.^^J
Perhaps you wanted \string\@\space for a printed @?^^J
I've ignored the character or group after @.}
\def\futureletnextat@{\futurelet\next\at@}
{\catcode`\@=\active
\lccode`\Z=`\@ \lowercase
{\gdef@{\expandafter\csname futureletnextatZ\endcsname}
\expandafter\gdef\csname atZ\endcsname
   {\ifcat\noexpand\next a\def\next{\csname atZZ\endcsname}\else
   \ifcat\noexpand\next0\def\next{\csname atZZ\endcsname}\else
    \def\next{\csname atZZZ\endcsname}\fi\fi\next}
\expandafter\gdef\csname atZZ\endcsname#1{\expandafter
   \ifx\csname #1Zat\endcsname\relax\def\next
     {\errhelp\expandafter=\csname athelpZ\endcsname
      \errmessage{Invalid use of \string@}}\else
       \def\next{\csname #1Zat\endcsname}\fi\next}
\expandafter\gdef\csname atZZZ\endcsname#1{\errhelp
    \expandafter=\csname athelpZ\endcsname
      \errmessage{Invalid use of \string@}}}}
\def\atdef@#1{\expandafter\def\csname #1@at\endcsname}
\newhelp\defahelp@{If you typed \string\define\space cs instead of
\string\define\string\cs\space^^J
I've substituted an inaccessible control sequence so that your^^J
definition will be completed without mixing me up too badly.^^J
If you typed \string\define{\string\cs} the inaccessible control sequence^^J
was defined to be \string\cs, and the rest of your^^J
definition appears as input.}
\newhelp\defbhelp@{I've ignored your definition, because it might^^J
conflict with other uses that are important to me.}
\def\define{\futurelet\next\define@}
\def\define@{\ifcat\noexpand\next\relax
  \def\next{\define@@}%
  \else\errhelp=\defahelp@
  \errmessage{\string\define\space must be followed by a control 
     sequence}\def\next{\def\garbage@}\fi\next}
\def\undefined@{}
\def\preloaded@{}    
\def\define@@#1{\ifx#1\relax\errhelp=\defbhelp@
   \errmessage{\string#1\space is already defined}\def\next{\def\garbage@}%
   \else\expandafter\ifx\csname\expandafter\eat@\string
         #1@\endcsname\undefined@\errhelp=\defbhelp@
   \errmessage{\string#1\space can't be defined}\def\next{\def\garbage@}%
   \else\expandafter\ifx\csname\expandafter\eat@\string#1\endcsname\relax
     \def\next{\def#1}\else\errhelp=\defbhelp@
     \errmessage{\string#1\space is already defined}\def\next{\def\garbage@}%
      \fi\fi\fi\next}
\def\famzero{\fam\z@}

\def\lim{\mathop{\famzero lim}}

\def\textfont@#1#2{\def#1{\relax\ifmmode
    \errmessage{Use \string#1\space only in text}\else#2\fi}}
\textfont@\rm\tenrm
\textfont@\it\tenit
\textfont@\sl\tensl
\textfont@\bf\tenbf
\textfont@\smc\tensmc
\let\ic@=\/
\def\/{\unskip\ic@}
\def\textfonti{\the\textfont1 }
\def\t#1#2{{\edef\next{\the\font}\textfonti\accent"7F \next#1#2}}
\let\B=\=
\let\D=\.
\def~{\unskip\nobreak\ \ignorespaces}
{\catcode`\@=\active
\gdef\@{\char'100 }}
\atdef@-{\leavevmode\futurelet\next\athyph@}
\def\athyph@{\ifx\next-\let\next=\athyph@@
  \else\let\next=\athyph@@@\fi\next}
\def\athyph@@@{\hbox{-}}
\def\athyph@@#1{\futurelet\next\athyph@@@@}
\def\athyph@@@@{\if\next-\def\next##1{\hbox{---}}\else
    \def\next{\hbox{--}}\fi\next}
\def\.{.\spacefactor=\@m}
\atdef@.{\null.}
\atdef@,{\null,}
\atdef@;{\null;}
\atdef@:{\null:}
\atdef@?{\null?}
\atdef@!{\null!}   
\def\srdr@{\thinspace}                     
\def\drsr@{\kern.02778em}
\def\sldl@{\kern.02778em}
\def\dlsl@{\thinspace}
\atdef@"{\unskip\futurelet\next\atqq@}
\def\atqq@{\ifx\next\Space@\def\next. {\atqq@@}\else
         \def\next.{\atqq@@}\fi\next.}
\def\atqq@@{\futurelet\next\atqq@@@}
\def\atqq@@@{\ifx\next`\def\next`{\atqql@}\else\def\next'{\atqqr@}\fi\next}
\def\atqql@{\futurelet\next\atqql@@}
\def\atqql@@{\ifx\next`\def\next`{\sldl@``}\else\def\next{\dlsl@`}\fi\next}
\def\atqqr@{\futurelet\next\atqqr@@}
\def\atqqr@@{\ifx\next'\def\next'{\srdr@''}\else\def\next{\drsr@'}\fi\next}

\def\textfontii{\the\textfont2 }
\def\{{\relax\ifmmode\lbrace\else
    {\textfontii f}\spacefactor=\@m\fi}
\def\}{\relax\ifmmode\rbrace\else
    \let\@sf=\empty\ifhmode\edef\@sf{\spacefactor=\the\spacefactor}\fi
      {\textfontii g}\@sf\relax\fi}   
\def\nonhmodeerr@#1{\errmessage
     {\string#1\space allowed only within text}}
\def\linebreak{\relax\ifhmode\unskip\break\else
    \nonhmodeerr@\linebreak\fi}
\def\allowlinebreak{\relax
   \ifhmode\allowbreak\else\nonhmodeerr@\allowlinebreak\fi}
\newskip\saveskip@
\def\nolinebreak{\relax\ifhmode\saveskip@=\lastskip\unskip
  \nobreak\ifdim\saveskip@>\z@\hskip\saveskip@\fi
   \else\nonhmodeerr@\nolinebreak\fi}
\def\newline{\relax\ifhmode\null\hfil\break
    \else\nonhmodeerr@\newline\fi}
\def\nonmathaerr@#1{\errmessage
     {\string#1\space is not allowed in display math mode}}
\def\nonmathberr@#1{\errmessage{\string#1\space is allowed only in math mode}}
\def\mathbreak{\relax\ifmmode\ifinner\break\else
   \nonmathaerr@\mathbreak\fi\else\nonmathberr@\mathbreak\fi}
\def\nomathbreak{\relax\ifmmode\ifinner\nobreak\else
    \nonmathaerr@\nomathbreak\fi\else\nonmathberr@\nomathbreak\fi}
\def\allowmathbreak{\relax\ifmmode\ifinner\allowbreak\else
     \nonmathaerr@\allowmathbreak\fi\else\nonmathberr@\allowmathbreak\fi}
\def\pagebreak{\relax\ifmmode
   \ifinner\errmessage{\string\pagebreak\space
     not allowed in non-display math mode}\else\postdisplaypenalty-\@M\fi
   \else\ifvmode\penalty-\@M\else\edef\spacefactor@
       {\spacefactor=\the\spacefactor}\vadjust{\penalty-\@M}\spacefactor@
        \relax\fi\fi}
\def\nopagebreak{\relax\ifmmode
     \ifinner\errmessage{\string\nopagebreak\space
    not allowed in non-display math mode}\else\postdisplaypenalty\@M\fi
    \else\ifvmode\nobreak\else\edef\spacefactor@
        {\spacefactor=\the\spacefactor}\vadjust{\penalty\@M}\spacefactor@
         \relax\fi\fi}
\def\newpage{\relax\ifvmode\vfill\penalty-\@M\else\nonvmodeerr@\newpage\fi}
\def\nonvmodeerr@#1{\errmessage
    {\string#1\space is allowed only between paragraphs}}
\def\smallpagebreak{\relax\ifvmode\smallbreak
      \else\nonvmodeerr@\smallpagebreak\fi}
\def\medpagebreak{\relax\ifvmode\medbreak
       \else\nonvmodeerr@\medpagebreak\fi}
\def\bigpagebreak{\relax\ifvmode\bigbreak
      \else\nonvmodeerr@\bigpagebreak\fi}
\newdimen\captionwidth@
\captionwidth@=\hsize
\advance\captionwidth@ by -1.5in
\def\caption#1{}
\def\topspace#1{\gdef\thespace@{#1}\ifvmode\def\next
    {\futurelet\next\topspace@}\else\def\next{\nonvmodeerr@\topspace}\fi\next}
\def\topspace@{\ifx\next\Space@\def\next. {\futurelet\next\topspace@@}\else
     \def\next.{\futurelet\next\topspace@@}\fi\next.}
\def\topspace@@{\ifx\next\caption\let\next\topspace@@@\else
    \let\next\topspace@@@@\fi\next}
 \def\topspace@@@@{\topinsert\vbox to 
       \thespace@{}\endinsert}
\def\topspace@@@\caption#1{\topinsert\vbox to
    \thespace@{}\nobreak
      \smallskip
    \setbox\z@=\hbox{\noindent\ignorespaces#1\unskip}%
   \ifdim\wd\z@>\captionwidth@
   \centerline{\vbox{\hsize=\captionwidth@\noindent\ignorespaces#1\unskip}}%
   \else\centerline{\box\z@}\fi\endinsert}
\def\midspace#1{\gdef\thespace@{#1}\ifvmode\def\next
    {\futurelet\next\midspace@}\else\def\next{\nonvmodeerr@\midspace}\fi\next}
\def\midspace@{\ifx\next\Space@\def\next. {\futurelet\next\midspace@@}\else
     \def\next.{\futurelet\next\midspace@@}\fi\next.}
\def\midspace@@{\ifx\next\caption\let\next\midspace@@@\else
    \let\next\midspace@@@@\fi\next}
 \def\midspace@@@@{\midinsert\vbox to 
       \thespace@{}\endinsert}
\def\midspace@@@\caption#1{\midinsert\vbox to
    \thespace@{}\nobreak
      \smallskip
      \setbox\z@=\hbox{\noindent\ignorespaces#1\unskip}%
      \ifdim\wd\z@>\captionwidth@
    \centerline{\vbox{\hsize=\captionwidth@\noindent\ignorespaces#1\unskip}}%
    \else\centerline{\box\z@}\fi\endinsert}
\mathchardef\prime@="0230
\def\prime{{{}\prime@{}}}
\def\prim@s{\prime@\futurelet\next\pr@m@s}

\def\,{\relax\ifmmode\mskip\thinmuskip\else\thinspace\fi}
\def\!{\relax\ifmmode\mskip-\thinmuskip\else\negthinspace\fi}
\def\frac#1#2{{#1\over#2}}

\def\:{\nobreak\hskip.1111em{:}\hskip.3333em plus .0555em\relax}
\def\intic@{\mathchoice{\hskip5\p@}{\hskip4\p@}{\hskip4\p@}{\hskip4\p@}}
\def\negintic@
 {\mathchoice{\hskip-5\p@}{\hskip-4\p@}{\hskip-4\p@}{\hskip-4\p@}}
\def\intkern@{\mathchoice{\!\!\!}{\!\!}{\!\!}{\!\!}}
\def\intdots@{\mathchoice{\cdots}{{\cdotp}\mkern1.5mu
    {\cdotp}\mkern1.5mu{\cdotp}}{{\cdotp}\mkern1mu{\cdotp}\mkern1mu
      {\cdotp}}{{\cdotp}\mkern1mu{\cdotp}\mkern1mu{\cdotp}}}
\newcount\intno@             
\def\iint{\intno@=\tw@\futurelet\next\ints@} 
\def\iiint{\intno@=\thr@@\futurelet\next\ints@}
\def\iiiint{\intno@=4 \futurelet\next\ints@}
\def\idotsint{\intno@=\z@\futurelet\next\ints@}
\def\ints@{\findlimits@\ints@@}
\newif\iflimtoken@
\newif\iflimits@
\def\findlimits@{\limtoken@false\limits@false\ifx\next\limits
 \limtoken@true\limits@true\else\ifx\next\nolimits\limtoken@true\limits@false
    \fi\fi}
\def\multintlimits@{\intop\ifnum\intno@=\z@\intdots@
  \else\intkern@\fi
    \ifnum\intno@>\tw@\intop\intkern@\fi
     \ifnum\intno@>\thr@@\intop\intkern@\fi\intop}
\def\multint@{\int\ifnum\intno@=\z@\intdots@\else\intkern@\fi
   \ifnum\intno@>\tw@\int\intkern@\fi
    \ifnum\intno@>\thr@@\int\intkern@\fi\int}
\def\ints@@{\iflimtoken@\def\ints@@@{\iflimits@
   \negintic@\mathop{\intic@\multintlimits@}\limits\else
    \multint@\nolimits\fi\eat@}\else
     \def\ints@@@{\multint@\nolimits}\fi\ints@@@}
\def\Sb{_\bgroup\vspace@
        \baselineskip=\fontdimen10 \scriptfont\tw@
        \advance\baselineskip by \fontdimen12 \scriptfont\tw@
        \lineskip=\thr@@\fontdimen8 \scriptfont\thr@@
        \lineskiplimit=\thr@@\fontdimen8 \scriptfont\thr@@
        \Let@\vbox\bgroup\halign\bgroup \hfil$\scriptstyle
            {##}$\hfil\cr}
\def\endSb{\crcr\egroup\egroup\egroup}
\def\Sp{^\bgroup\vspace@
        \baselineskip=\fontdimen10 \scriptfont\tw@
        \advance\baselineskip by \fontdimen12 \scriptfont\tw@
        \lineskip=\thr@@\fontdimen8 \scriptfont\thr@@
        \lineskiplimit=\thr@@\fontdimen8 \scriptfont\thr@@
        \Let@\vbox\bgroup\halign\bgroup \hfil$\scriptstyle
            {##}$\hfil\cr}
\def\endSp{\crcr\egroup\egroup\egroup}
\def\Let@{\relax\iffalse{\fi\let\\=\cr\iffalse}\fi}
\def\vspace@{\def\vspace##1{\noalign{\vskip##1 }}}
\def\aligned{\,\vcenter\bgroup\vspace@\Let@\openup\jot\m@th\ialign
  \bgroup \strut\hfil$\displaystyle{##}$&$\displaystyle{{}##}$\hfil\crcr}
\def\endaligned{\crcr\egroup\egroup}
\def\matrix{\,\vcenter\bgroup\Let@\vspace@
    \normalbaselines
  \m@th\ialign\bgroup\hfil$##$\hfil&&\quad\hfil$##$\hfil\crcr
    \mathstrut\crcr\noalign{\kern-\baselineskip}}
\def\endmatrix{\crcr\mathstrut\crcr\noalign{\kern-\baselineskip}\egroup
                \egroup\,}
\newtoks\hashtoks@
\hashtoks@={#}
\def\format{\crcr\egroup\iffalse{\fi\ifnum`}=0 \fi\format@}
\def\format@#1\\{\def\preamble@{#1}%
  \def\c{\hfil$\the\hashtoks@$\hfil}%
  \def\r{\hfil$\the\hashtoks@$}%
  \def\l{$\the\hashtoks@$\hfil}%
  \setbox\z@=\hbox{\xdef\Preamble@{\preamble@}}\ifnum`{=0 \fi\iffalse}\fi
   \ialign\bgroup\span\Preamble@\crcr}

\def\cases{\left\{\,\vcenter\bgroup\vspace@
     \normalbaselines\openup\jot\m@th
       \Let@\ialign\bgroup$##$\hfil&\quad$##$\hfil\crcr
      \mathstrut\crcr\noalign{\kern-\baselineskip}}

\newif\iftagsleft@
\tagsleft@true
\def\TagsOnRight{\global\tagsleft@false}
\def\tag#1$${\iftagsleft@\leqno\else\eqno\fi
 \hbox{\def\pagebreak{\global\postdisplaypenalty-\@M}%
 \def\nopagebreak{\global\postdisplaypenalty\@M}\rm(#1\unskip)}%
  $$\postdisplaypenalty\z@\ignorespaces}
\interdisplaylinepenalty=\@M
\def\allowdisplaybreak@{\def\allowdisplaybreak{\noalign{\allowbreak}}}
\def\displaybreak@{\def\displaybreak{\noalign{\break}}}
\def\align#1\endalign{\def\tag{&}\vspace@\allowdisplaybreak@\displaybreak@
  \iftagsleft@\lalign@#1\endalign\else
   \ralign@#1\endalign\fi}
\def\ralign@#1\endalign{\displ@y\Let@\tabskip\centering\halign to\displaywidth
     {\hfil$\displaystyle{##}$\tabskip=\z@&$\displaystyle{{}##}$\hfil
       \tabskip=\centering&\llap{\hbox{(\rm##\unskip)}}\tabskip\z@\crcr
             #1\crcr}}
\def\lalign@
 #1\endalign{\displ@y\Let@\tabskip\centering\halign to \displaywidth
   {\hfil$\displaystyle{##}$\tabskip=\z@&$\displaystyle{{}##}$\hfil
   \tabskip=\centering&\kern-\displaywidth
        \rlap{\hbox{(\rm##\unskip)}}\tabskip=\displaywidth\crcr
               #1\crcr}}
\def\overrightarrow{\mathpalette\overrightarrow@}
\def\overrightarrow@#1#2{\vbox{\ialign{$##$\cr
    #1{-}\mkern-6mu\cleaders\hbox{$#1\mkern-2mu{-}\mkern-2mu$}\hfill
     \mkern-6mu{\to}\cr
     \noalign{\kern -1\p@\nointerlineskip}
     \hfil#1#2\hfil\cr}}}
\def\overleftarrow{\mathpalette\overleftarrow@}
\def\overleftarrow@#1#2{\vbox{\ialign{$##$\cr
     #1{\leftarrow}\mkern-6mu\cleaders\hbox{$#1\mkern-2mu{-}\mkern-2mu$}\hfill
      \mkern-6mu{-}\cr
     \noalign{\kern -1\p@\nointerlineskip}
     \hfil#1#2\hfil\cr}}}
\def\overleftrightarrow{\mathpalette\overleftrightarrow@}
\def\overleftrightarrow@#1#2{\vbox{\ialign{$##$\cr
     #1{\leftarrow}\mkern-6mu\cleaders\hbox{$#1\mkern-2mu{-}\mkern-2mu$}\hfill
       \mkern-6mu{\to}\cr
    \noalign{\kern -1\p@\nointerlineskip}
      \hfil#1#2\hfil\cr}}}
\def\underrightarrow{\mathpalette\underrightarrow@}
\def\underrightarrow@#1#2{\vtop{\ialign{$##$\cr
    \hfil#1#2\hfil\cr
     \noalign{\kern -1\p@\nointerlineskip}
    #1{-}\mkern-6mu\cleaders\hbox{$#1\mkern-2mu{-}\mkern-2mu$}\hfill
     \mkern-6mu{\to}\cr}}}
\def\underleftarrow{\mathpalette\underleftarrow@}
\def\underleftarrow@#1#2{\vtop{\ialign{$##$\cr
     \hfil#1#2\hfil\cr
     \noalign{\kern -1\p@\nointerlineskip}
     #1{\leftarrow}\mkern-6mu\cleaders\hbox{$#1\mkern-2mu{-}\mkern-2mu$}\hfill
      \mkern-6mu{-}\cr}}}
\def\underleftrightarrow{\mathpalette\underleftrightarrow@}
\def\underleftrightarrow@#1#2{\vtop{\ialign{$##$\cr
      \hfil#1#2\hfil\cr
    \noalign{\kern -1\p@\nointerlineskip}
     #1{\leftarrow}\mkern-6mu\cleaders\hbox{$#1\mkern-2mu{-}\mkern-2mu$}\hfill
       \mkern-6mu{\to}\cr}}}
\def\sqrt#1{\radical"270370 {#1}}
\def\dots{\relax\ifmmode\let\next=\ldots\else\let\next=\tdots@\fi\next}
\def\tdots@{\unskip\ \tdots@@}
\def\tdots@@{\futurelet\next\tdots@@@}
\def\tdots@@@{$\mathinner{\ldotp\ldotp\ldotp}\,
   \ifx\next,$\else
   \ifx\next.\,$\else
   \ifx\next;\,$\else
   \ifx\next:\,$\else
   \ifx\next?\,$\else
   \ifx\next!\,$\else
   $ \fi\fi\fi\fi\fi\fi}
\def\text{\relax\ifmmode\let\next=\text@\else\let\next=\text@@\fi\next}
\def\text@@#1{\hbox{#1}}
\def\text@#1{\mathchoice
 {\hbox{\everymath{\displaystyle}\def\textfonti{\the\textfont1 }%
    \def\textfontii{\the\textfont2 }\textdef@@ T#1}}
 {\hbox{\everymath{\textstyle}\def\textfonti{\the\textfont1 }%
    \def\textfontii{\the\textfont2 }\textdef@@ T#1}}
 {\hbox{\everymath{\scriptstyle}\def\textfonti{\the\scriptfont1 }%
   \def\textfontii{\the\scriptfont2 }\textdef@@ S\rm#1}}
 {\hbox{\everymath{\scriptscriptstyle}\def\textfonti{\the\scriptscriptfont1 }%
   \def\textfontii{\the\scriptscriptfont2 }\textdef@@ s\rm#1}}}
\def\textdef@@#1{\textdef@#1\rm \textdef@#1\bf
   \textdef@#1\sl \textdef@#1\it}

\def\textdef@#1#2{\def\next{\csname\expandafter\eat@\string#2fam\endcsname}%
\if S#1\edef#2{\the\scriptfont\next\relax}%
 \else\if s#1\edef#2{\the\scriptscriptfont\next\relax}%
 \else\edef#2{\the\textfont\next\relax}\fi\fi}
\scriptfont\itfam=\tenit \scriptscriptfont\itfam=\tenit
\scriptfont\slfam=\tensl \scriptscriptfont\slfam=\tensl
\mathcode`\0="0030
\mathcode`\1="0031
\mathcode`\2="0032
\mathcode`\3="0033
\mathcode`\4="0034
\mathcode`\5="0035
\mathcode`\6="0036
\mathcode`\7="0037
\mathcode`\8="0038
\mathcode`\9="0039
\def\Cal{\relax\ifmmode\let\next=\Cal@\else
     \def\next{\errmessage{Use \string\Cal\space only in math mode}}\fi\next}
\def\Cal@#1{{\fam2 #1}}
\def\bold{\relax\ifmmode\let\next=\bold@\else
   \def\next{\errmessage{Use \string\bold\space only in math
      mode}}\fi\next}\def\bold@#1{{\fam\bffam #1}}
\mathchardef\Gamma="0000
\mathchardef\Delta="0001
\mathchardef\Theta="0002
\mathchardef\Lambda="0003
\mathchardef\Xi="0004
\mathchardef\Pi="0005
\mathchardef\Sigma="0006
\mathchardef\Upsilon="0007
\mathchardef\Phi="0008
\mathchardef\Psi="0009
\mathchardef\Omega="000A
\mathchardef\varGamma="0100
\mathchardef\varDelta="0101
\mathchardef\varTheta="0102
\mathchardef\varLambda="0103
\mathchardef\varXi="0104
\mathchardef\varPi="0105
\mathchardef\varSigma="0106
\mathchardef\varUpsilon="0107
\mathchardef\varPhi="0108
\mathchardef\varPsi="0109
\mathchardef\varOmega="010A
\font\dummyft@=dummy
\fontdimen1 \dummyft@=\z@
\fontdimen2 \dummyft@=\z@
\fontdimen3 \dummyft@=\z@
\fontdimen4 \dummyft@=\z@
\fontdimen5 \dummyft@=\z@
\fontdimen6 \dummyft@=\z@
\fontdimen7 \dummyft@=\z@
\fontdimen8 \dummyft@=\z@
\fontdimen9 \dummyft@=\z@
\fontdimen10 \dummyft@=\z@
\fontdimen11 \dummyft@=\z@
\fontdimen12 \dummyft@=\z@
\fontdimen13 \dummyft@=\z@
\fontdimen14 \dummyft@=\z@
\fontdimen15 \dummyft@=\z@
\fontdimen16 \dummyft@=\z@
\fontdimen17 \dummyft@=\z@
\fontdimen18 \dummyft@=\z@
\fontdimen19 \dummyft@=\z@
\fontdimen20 \dummyft@=\z@
\fontdimen21 \dummyft@=\z@
\fontdimen22 \dummyft@=\z@
\def\fontlist@{\\{\tenrm}\\{\sevenrm}\\{\fiverm}\\{\teni}\\{\seveni}%
 \\{\fivei}\\{\tensy}\\{\sevensy}\\{\fivesy}\\{\tenex}\\{\tenbf}\\{\sevenbf}%
 \\{\fivebf}\\{\tensl}\\{\tenit}\\{\tensmc}}
\def\dodummy@{{\def\\##1{\global\let##1=\dummyft@}\fontlist@}}
\newif\ifsyntax@
\newcount\countxviii@
\def\newtoks@{\alloc@5\toks\toksdef\@cclvi}
\def\nopages@{\output={\setbox\z@=\box\@cclv \deadcycles=\z@}\newtoks@\output}
\def\syntax{\syntax@true\dodummy@\countxviii@=\count18
\loop \ifnum\countxviii@ > \z@ \textfont\countxviii@=\dummyft@
   \scriptfont\countxviii@=\dummyft@ \scriptscriptfont\countxviii@=\dummyft@
     \advance\countxviii@ by-\@ne\repeat
\dummyft@\tracinglostchars=\z@
  \nopages@\frenchspacing\hbadness=\@M}
\def\magstep#1{\ifcase#1 1000\or
 1200\or 1440\or 1728\or 2074\or 2488\or 
 \errmessage{\string\magstep\space only works up to 5}\fi\relax}
{\lccode`\2=`\p \lccode`\3=`\t 
 \lowercase{\gdef\tru@#123{#1truept}}}

\def\scaletype#1{\mag=#1\relax
 \hsize=\expandafter\tru@\the\hsize
 \vsize=\expandafter\tru@\the\vsize
 \dimen\footins=\expandafter\tru@\the\dimen\footins}

\def\scalefont#1#2\andcallit#3{\edef\font@{\the\font}#1\font#3=
  \fontname\font\space scaled #2\relax\font@}
\def\Mag@#1#2{\ifdim#1<1pt\multiply#1 #2\relax\divide#1 1000 \else
  \ifdim#1<10pt\divide#1 10 \multiply#1 #2\relax\divide#1 100\else
  \divide#1 100 \multiply#1 #2\relax\divide#1 10 \fi\fi}
\def\scalelinespacing#1{\Mag@\baselineskip{#1}\Mag@\lineskip{#1}%
  \Mag@\lineskiplimit{#1}}
\def\wlog#1{\immediate\write-1{#1}}
\catcode`\@=\active

%%%%%%%%%%%%%%%%%%%%%%%%%%%%%%%%%%%%%%%%%%%%%%%%%%%%%%%%%%%%%
%2. input: our macros

%paper format:
\magnification 1200 %optional
\hsize=16.6truecm%15.6truecm
\vsize=22.6truecm %%%cancel optionally
\advance\hoffset by -.5truecm
\advance\baselineskip by 6pt
\nopagenumbers
\footline={\hss}

\TagsOnRight

%fonts
\font\sixrm=cmr6        \font\eightrm=cmr8
        
\font\gross=cmcsc10     \font\gr=cmcsc10      
     
    1
 2      
\font\Titel=cmr10 scaled \magstep 2

\hyphenation{
Meas-ure-ment Meas-uring 
meas-ure-ment meas-ure meas-uring
pre-meas-ure-ment pre-meas-ure pre-meas-uring}

%long phrases
\def\sy{system}           
\def\qm{quantum mechanics}

\def\me{meas\-ure}                    \def\mt{\me{}ment} 

\def\pov{{\gross pov} meas\-ure}      \def\pv{{\gross pv} \me}

\def\op{operator}                    \def\ops{operators}
\def\ob{observable}                  \def\obs{observables}
\def\sad{self-adjoint}               \def\sop{\sad\ \op}

%operators:
           %projector on state
\def\bo#1{{\bold #1}}                    %for use in math mode
             %norm   
\def\ip#1#2{\left\langle\,#1\,|\,#2\,\right\rangle} %inner product
\def\kb#1#2{|#1\,\rangle\langle\,#2|}    %Dirac operators

\def\tr#1{\text{tr}\bigl[#1\bigr]}       %trace
\def\fii{\varphi}

%math objects
\def\hi{{\Cal H}}                     
              
\def\s{$\Cal S$}         \def\a{$\Cal A$}      \def\sa{$\Cal S + \Cal A$}
\def\os{(\Omega,\Sigma)}

%%%%%%%%%%%%%%%%%%%%%%%%%%%%%%%%%%%%%%%%%%%%%%%%%%%%%%%%%%%%%%%%
%3. text

\headline{\centerline{\sixrm Insolubility of the Quantum Measurement
Problem for Unsharp Observables\hfill \folio}}

\topinsert
\noindent{\sixrm Studies in History and Philosophy of
Modern Physics -- to appear, 1996}
\endinsert

\title{\Titel Insolubility of the Quantum Measurement Problem}\\ 
{\Titel for Unsharp Observables}
\endtitle

%\vskip .7truecm
\centerline{{\gross Paul Busch}}
%\vskip -5pt
\centerline{\eightrm Department of Applied Mathematics, University of
Hull}
%\vskip -5pt
\centerline{\eightrm Hull, HU6 7RX, UK}
%\vskip -5pt
\centerline{\eightrm E-mail: p.busch\@maths.hull.ac.uk}
\vskip 0pt
\centerline{\gross Abner Shimony}
%\vskip -5pt
\centerline{\eightrm Departments of Physics and Philosophy, Boston
University}
%\vskip -5pt
\centerline{\eightrm Boston, MA 02215, USA}

\vskip .4truecm
\centerline{\gross Abstract}
%\vskip 3pt

\noindent
 The quantum mechanical measurement problem is the difficulty of dealing
with the indefiniteness of the pointer observable at the conclusion of a
meas\-urement process governed by unitary quantum dynamics. There has been
hope to solve this problem by eliminating idealizations from the
characterization of meas\-urement. We state and prove two `insolubility
theorems' that disappoint this hope. In both the initial state of the
apparatus is taken to be mixed rather than pure, and the correlation of the
object observable and the pointer observable is allowed to be imperfect.
In the {\it insolubility theorem for sharp observables}, which is only a modest
extension of previous results, the object observable is taken to be an
arbitrary projection valued meas\-ure. In the {\it insolubility theorem for
unsharp observables}, which is essentially new, the object observable is
taken to be a positive operator valued meas\-ure. Both theorems show that
the meas\-urement problem is not the consequence of neglecting the
ever-present imperfections of actual meas\-urements.

\vskip 3pt

%\noindent {\bf PACS number:} 03.65.Bz.

\vfil
%\noindent$\underline{\text{\hskip 4cm}}$\hfill
%\vskip -5pt
%\noindent {$^\dagger$} {\eightrm Forthcoming address: Department of
%Applied Mathematics, University of Hull, Hull HU6 7RX, UK}
\eject

\subheading{1. Introduction}

\noindent
  The quantum mechanical \mt\ problem consists of the difficulty of
reconciling the occurrence of definite, objective `pointer' readings with
the unitarity and thus linearity of the time evolution of quantum
states. The problem is posed simply if one considers the meas\-urement
of a 
discrete sharp observable $A=\sum_i a_i\kb{\fii_i}{\fii_i}$ and requires
the following {\it calibration condition}: 
if the object \sy\ \s\ is in an eigenstate of $A$, say $\fii_k$, then the
state of the apparatus \a\ after the interaction with the object is an
eigenstate of the pointer \ob\ $Z$ associated with a pointer reading $z_k$
indicating that the value of $A$ was $a_k$. Hence if the initial state of
\sa\ is $\fii_k\otimes\phi$ and the \mt\ coupling is described by a unitary
\op\ $U$, then the final state $U(\fii_k\otimes\phi)$ must be an eigenstate
of the observable $I\otimes Z$. But then the linearity of $U$ entails that
for any state $\fii$ of \s\ which is not an eigenstate of $A$, the
resulting state of \sa\ is a superposition of
eigenstates of $I\otimes Z$. Certainly this does not correspond to a
situation where the pointer observable is objectified.

It has sometimes been suggested that the quantum mechanical meas\-urement
problem is spurious, resulting from the excessively idealized
characterization of the \mt\ process. One might conjecture that a more
realistic characterization, taking into account the practical impossibility
of preparing the macroscopic apparatus in a pure quantum state and
acknowledging the possibility of imperfect correlation between the object
observable and the pointer observable, would eliminate the \mt\ problem by
assuring that the pointer observable is objectified. A series of papers,
initiated by Wigner$^1$ and continued by various authors, including
d'Espagnat$^2$, Fine$^3$, and Shimony$^4$, 
has thrown grave doubts on this conjecture in the important case when the
object observable is a self-adjoint operator (equivalently, when it is
represented by a spectral meas\-ure).
In Sec.\ 2, the strongest result in this series, that of Ref.\ 4,
is extended by taking the observable to be an arbitrary projection valued
meas\-ure. We shall call the extended theorem  
the {\it insolubility theorem for sharp observables}.

The primary purpose of this paper is to consider the consequences of
removing one more idealization in the usual treatments of \mt: viz., to
replace the assumption that the object observables are sharp by the more
realistic assumption that they are (or may be) unsharp. In Section 3 the 
\mt\ process of unsharp observables will be characterized mathematically.
We shall then demonstrate the main result of this paper: the 
{\it insolubility theorem for unsharp observables}, where an unsharp observable
is represented as a positive operator valued \me.

\subheading{2. Insolubility theorem for sharp \obs}

\noindent
 In standard formulations of \qm, the {\it pure states} of a \sy\ are
identified with rays (one-dimensional subspaces) of a Hilbert space $\hi$,
or equivalently, with projection operators onto rays. General {\it states}
(including both {\it mixed} and {\it pure} states) are represented by
density or state operators on $\hi$. For any unit vector $\fii\in\hi$, the
corresponding projection shall be denoted $P_\fii$.  {\it Observables} of
the system are identified with \sad\ \ops\ on $\hi$.  (We use the terms
{\it states} and {\it observables} both for the physical entities and the
mathematical objects representing them.)  By the classical spectral theorem
of von Neumann, a \sop\ $A$ can be expressed in terms of a family of
projection operators. In modern mathematical locution, with any \sop\ $A$
there is associated a unique projection valued \me, its spectral \me\
$E^A$.  This assertion makes it possible to recover the expectation values
$\langle A\rangle_\fii$ %($\fii$ being a unit vector in $\hi$) as averages
over the probability \me{}s $X\mapsto \ip{\fii}{E^A(X)\fii}$ (where $X$
runs through the (Borel) subsets of $\bo R$).

A first extension of the set of observables is obtained by admitting more
general {\it projection valued} ({\gr pv}) {\it \me{}s}, 
defined with respect to a
meas\-urable space $\os$, where $\Omega$ is a set and $\Sigma$ is a
$\sigma-$algebra of subsets of $\Omega$: i.e., a \pv\ is a map $E$ from
$\Sigma$ into the lattice of projections such that $E(\emptyset)=O$,
$E(\Omega)=I$ ($O,I$ denoting the null and unit operators in $\hi$,
respectively), and $E(\cup_iX_i)=\sum_i E(X_i)$ for any countable
collection of mutually disjoint sets $X_i\in\Sigma$.  These conditions
ensure that for any state operator $T$ of \s, the map $X\mapsto
\tr{TE(X)}=: p_T^E(X)$ is a probability \me\ on $\os$. Since $\Omega$
concerns the set of possible values of the physical quantity represented
by $E$, $\os$ is called the {\it value space} of that observable. The case of a
spectral \me\ is recovered by choosing the real line for $\Omega$ and the
Borel sets for $\Sigma$. The introduction of more general value spaces
$\os$ and of observables as \pv{}s on them proves convenient for a variety
of purposes, such as, for example, the description of joint \mt{}s of
several commuting \obs. Henceforth, we shall use the term {\it sharp
observable} for an observable represented by a general \pv,$^5$ 
in anticipation of a further 
generalization in Section 3 to unsharp observables.

In the remainder of this paper we shall take $\hi$ and $\hi_a$ to be 
the Hilbert spaces associated respectively with the object system \s\ and
the apparatus \a. In the spirit of Ref.\ 5 we shall call the
quadruple $\langle \hi_a,Z,T_a,U\rangle$ a {\it \mt\ scheme} for 
\s\ if $Z$ is the sharp
pointer \ob\ of \a, $T_a$ an initial state of \a, and $U$ a unitary \op\ on
$\hi\otimes\hi_a$, the tensor product space associated with \sa. 
Note that the concept of a meas\-urement scheme does not explicitly refer
to an observable of the object \s, but the following useful concepts do so.

%The following definitions of distinguishability and discrimination are 
%based on criteria introduced by Fine$^3$ who used them as a
%characterization of a \mt\ of a sharp \ob\ represented by a \sop.

\proclaim{Definition 1} Let $E$ be a sharp \ob\ of a \sy\ \s.
Two state \ops\ $T$ and $T'$  are $E-$distinguishable if and only if 
$p_T^E\ne p_{T'}^E$.
\endproclaim

\proclaim{Definition 2} A \mt\ scheme $\langle \hi_a,Z,T_a,U\rangle$ for
\s\ is a discrimination of the sharp \ob\ $E$ if and only if the
$E-$distinguishability of any two states $T,T'$ of \s\ implies the
$I\otimes Z-$distinguishability of $U\big(T\otimes T_a\big)U^{-1}$ and
$U\big(T'\otimes T_a\big)U^{-1}$.
 \endproclaim

\proclaim{Definition 3} Given a sharp \ob\ $E$, a \mt\ scheme $\langle
\hi_a,Z,T_a,U\rangle$ is an $E-$\mt\ if and only if
the \pv\ $Z$ is defined on the value space of $E$ and the probability
reproducibility condition is fulfilled: for all states $T$ of \s\ and all
$X\in\Sigma$, 
$$
p_T^E(X)\ =\ p_{U(T\otimes T_a)U^{-1}}^{I\otimes Z}(X)\tag 1
$$
\endproclaim

Definitions 1 and 2 were essentially given by Fine$^3$, but with $E$
restricted to a spectral meas\-ure (equivalently, to a self-adjoint
operator). We have departed from Fine's terminology, however, by using the
term $E-${\it discrimination} in Definition 2 when he uses the term
{\it meas\-urement}. $E-$discrimination as defined seems to us the weakest,
hence the most general, condition connecting a meas\-urement scheme with an
observable, while intuitively meas\-urement is a more stringent 
and special
case of discrimination. Furthermore, Definition 3 singles out a unique
observable as the one meas\-ured by a given meas\-urement scheme, whereas
Definition 2 does not do so.

In order to formulate the insolubility theorem for sharp observables in a
way that applies to general \pv{}s, including continuous observables, we
need the concept of a {\it reading scale} for the pointer observable $Z$
as a partition of its value space $\os$, that is, a sequence of subsets 
$(X_i)\subseteq \Sigma$ such that $X_i\cap X_j=\emptyset$ if $i\ne j$ and
$\cup X_i=\Omega$.
Furthermore, it is necessary to exclude trivial
\pv{}s from the discussion: a \pv\ on $\os$ is {\it trivial} if there is a
point $\omega\in\Omega$ such that $E\big(\{\omega\}\big)=I$. Equivalently,
a \pv\ is non-trivial if there are at least two $E-$distinguishable states.

\proclaim{Insolubility Theorem for Sharp Observables} Let $E$ be a
non-trivial \pv. There is no $E-$discrimination (and hence no $E-$\mt)
$\langle\hi_a,Z,T_a,U\rangle$ such that for all initial \s\ states $T$ the
\sa\ states $U\big(T\otimes T_a\big)U^{-1}$ are mixtures of eigenstates of
the projections $I\otimes Z(X_i)$ for a given reading scale $(X_i)$.
 \endproclaim

\noindent
This theorem was essentially formulated by Fine$^3$ for the important case
in which $E$ is a discrete spectral meas\-ure. The proof which he gives,
however, does not establish the theorem as he asserted it. As Brown$^6$
points out, Fine's proof does establish a weaker theorem. This matter is
discussed in Fine$^7$ and Shimony$^8$ (and see also the recent discussion
of Stein$^9$).
Our proof of the insolubility
theorem will follow (but with a generalization to an arbitrary \pv)
the procedure of Shimony$^{4}$, who begins by proving an
auxiliary theorem. 

\proclaim{Inheritance of Superpositions Theorem} Hypotheses:
\item{\ \, (i)} $\fii_1,\fii_2$ are normalized orthogonal vectors of $\hi$;
$\{F_m\}$ is a (finite or countable) family of mutually orthogonal 
projections of $\hi_a$;
$U$ is a unitary operator on $\hi\otimes\hi_a$, and
$T_a$ is a state \op\ on $\hi_a$.

\item{\ (ii)} For some value of n
$$
\text{\rm tr}
\big[I\otimes F_n\, U\big(P_{\fii_1}\otimes T_a\big)U^{-1}\big]\ \ne\ 
\text{\rm tr}
\big[I\otimes F_n\, U\big(P_{\fii_2}\otimes T_a\big)U^{-1}\big]\tag 2
$$

\item{(iii)} There exist orthonormal sets $\{\xi^1_{nr}\}$, $\{\xi^2_{nr}\}$
and sets of positive numbers $\{b^1_{nr}\}$, $\{b^2_{nr}\}$ such that 
$\xi^j_{nr}\,\in\, \hi\otimes F_n\bigl(\hi_a\bigr)$ for $j=1,2$ and
$$
U\big(P_{\fii_j}\otimes T_a\big)U^{-1}\ =\ \sum_{n,r}b^j_{nr}\,P_{\xi^j_{nr}}
\tag 3
$$

\noindent
Conclusion: 
If $\fii\in\hi$ is defined as $c_1\fii_1+c_2\fii_2$, with both $c_1$
and $c_2$ nonzero, then there exists no orthonormal set $\{\psi_{nr}\}$
with $\psi_{nr}\in\hi\otimes F_n\bigl(\Cal H_a\bigr)$ and no positive 
coefficients $\{b_{nr}\}$ 
such that $\sum_{n,r}b_{nr}=1$ and 
$$
U\big(P_{\fii}\otimes T_a\big)U^{-1}\ =\ \sum_{n,r}b_{nr}\,P_{\psi_{nr}}
\tag 4
$$
\endproclaim

We do not reproduce the lengthy proof of this theorem here.$^4$
Hypothesis (ii) is a minimal requirement of information transfer in the
spirit of the concept of discrimination.
Condition (iii) corresponds to the idea that a \mt\ coupling
should lead to a final state which is a mixture of pointer eigenstates.
Then the conclusion states that these two conditions cannot be reconciled
with each other. It is not difficult to realize (4) if (ii) is given up.
Let $U$ be of the form $V\otimes V_a$, where $V,V_a$ are unitary operators
and $V_a$ commutes with all $F_n$. Further let $T_a$ be a mixture of
$F_n-$eigenstates. Then $U\bigl(P_\fii\otimes T_a\bigr)U^{-1}=
VP_\fii V^{-1}\otimes V_aT_aV_a^{-1}$ is a mixture of $I\otimes
F_n-$eigenstates, in fulfilment of (4). But the probabilities $\tr{I\otimes
F_n\,U\bigl(P_\fii\otimes T_a\bigr)U^{-1}}=\tr{F_n\,V_aT_aV_a^{-1}}=
\tr{F_nT_a}$ are independent of $\fii$ so that condition (ii) is 
violated.

We are now ready to prove the insolubility theorem for sharp observables.  
Suppose
$\fii_1,\fii_2$ are two normalized eigenvectors of the projections $E(X)$,
$E(Y)$, respectively, with disjoint sets $X,Y$: i.e., $E(X)\fii_1=\fii_1$,
$E(Y)\fii_2=\fii_2$.  Then $\fii_1$ and $\fii_2$ are mutually orthogonal
and indeed $E-$distinguishable. Suppose $U\big(P_{\fii_1}\otimes
T_a\big)U^{-1}$ and $U\big(P_{\fii_2}\otimes T_a\big)U^{-1}$ are $I\otimes
Z-$distinguishable and are both expressible as mixtures of eigenstates of
the projections $I\otimes Z(X_i)$.  Then the hypotheses of the inheritance
of superpositions theorem  are
satisfied. Therefore $U\big(P_\fii\otimes T_a\big)U^{-1}$ is not a mixture
of eigenstates of the $I\otimes Z(X_i)$, if $\fii$ is the superposition
$c_1\fii_1+c_2\fii_2$ with nonzero coefficients $c_1,c_2$. On the other
hand, if vectors $\fii_1,\fii_2$ with the assumed properties do not exist,
then evidently the claim of the insolubility theorem would also hold.

As long as we restrict our attention to sharp observables, 
it is hard to envisage further
strengthenings of this no-go verdict in the sense of extensions to more
realistic meas\-urement situations.

\subheading{3. Insolubility theorem for unsharp observables}

\noindent
 So far the inevitable imperfections of actual \mt{}s have been
acknowledged by weakening the correlation between sharp object observables
and pointer observables. Indeed, the concept of an $E-$discrimination
in Definition 2 is an extreme expression of such weakening. Another way of
being realistic about meas\-urements and acknowledging their imperfections
is to take the object observables themselves to be unsharp.$^{10}$
Therefore we consider now the case of a general \ob, 
represented as a positive operator valued \me\  
$E$ on a measurable space $\bigl(\Omega,\Sigma\bigr)$, its value space.

The map $E:X\mapsto E(X)$  is a  {\it positive operator valued} 
({\gross pov}) {\it measure} if the following conditions are
satisfied:  for each $X\in\Sigma$, $E(X)$ 
is an operator on the underlying Hilbert space
$\hi$ such that $O\le E(X)\le I$ [the ordering being in the sense of
expectation values; i.e., $A\le B$ if and only if $\ip\fii{A\fii}\le
\ip\fii{B\fii}$ for all $\fii\in\hi$]; moreover, $E(\emptyset)=O$ and
$E(\Omega)=I$, and $E(\cup_i X_i)=\sum_i E(X_i)$ for any countable
pairwise disjoint family $\{X_i\}\subseteq \Sigma$. These properties of $E$
ensure that  the map $p_T^E: X\mapsto
p_T^E(X):=\tr{TE(X)}$ is a probability \me\ for each state $T$. The special
case of \pv{}s is recovered if the additional property of idempotency,
$E(X)^2=E(X)$, is stipulated. 
%Obviously, this requirement is redundant from
%a probabilistic point of view. 

Note that the idempotency condition can be written as
$E(X)E(\Omega\setminus X)=O$, where $\Omega\setminus X$ is the complement
of $X$ so that $E(\Omega\setminus X)=I-E(X)$. It follows immediately
that a \pov\ is a sharp \ob\ 
if and only if for any two disjoint sets $X,Y$ the operators $E(X)$,
$E(Y)$ satisfy $E(X)E(Y)=O$. Such projections are orthogonal to each other
in the sense that their ranges are mutually orthogonal subspaces.
By contrast, for all other \pov{}s there will be sets $X$ such that
$E(X)$ and $E(\Omega\setminus X)$ are non-orthogonal. Such \pov{}s shall be
called {\it unsharp observables}.$^{11}$

We now can generalize the concepts given in Definitions 1,2 and 3. We use
the term \mt\ scheme for $\langle\hi_a,Z,T_a,U\rangle$ as given in Sec.\ 
2. Note that the pointer $Z$ is a \pv\ even though we are generalizing the
notion of an observable for the object system.

\proclaim{Definition $1'$} Let $E$ be a \pov\ associated with a \sy\ \s.
Two state \ops\ $T$ and $T'$  are $E-$distinguishable if and only if 
$p_T^E\ne p_{T'}^E$.
\endproclaim

\proclaim{Definition $2'$} A \mt\ scheme $\langle \hi_a,Z,T_a,U\rangle$ for
\s\ is a discrimination of the \pov\ $E$ if and only if the
$E-$distinguishability of any two states $T,T'$ of \s\ implies the
$I\otimes Z-$distinguishability of $U\big(T\otimes T_a\big)U^{-1}$ and
$U\big(T'\otimes T_a\big)U^{-1}$.
 \endproclaim

\proclaim{Definition $3'$} Given a \pov\ $E$, a \mt\ scheme $\langle
\hi_a,Z,T_a,U\rangle$ is an $E-$\mt\ if and only if
the \pv\ $Z$ is defined on the value space of $E$ and the probability
reproducibility condition is fulfilled: for all states $T$ of \s\ and all
$X\in\Sigma$, 
$$
p_T^E(X)\ =\ p_{U(T\otimes T_a)U^{-1}}^{I\otimes Z}(X)\tag 5
$$
\endproclaim
\noindent
 As mentioned in Section 2, one can read this definition in the opposite
direction in the sense that any \mt\ scheme $\langle\hi_a,Z,T_a,U\rangle$
induces, via Eq.\ (4), a unique \ob\ $E$ as the \me{}d one. As an example,
the probabilities $\tr{I\otimes F_n\,U(P_\fii\otimes T_a)U^{-1}}$ occurring
in Eq.\ (2) can be expressed as $\ip\fii{E_n\fii}$, defining thereby a set
of positive operators $E_n$ which constitute the \me{}d \ob\ given a
pointer \ob\ $Z$ constituted by the projections $F_n$.

In the following only non-trivial \obs\ $E$ will be
considered. $E$ is {\it trivial} if and only if it is of the form $E:X\mapsto
E(X):=\lambda(X)I$ for some probability \me\ $\lambda$. Equivalently,
$E$ is {\it non-trivial} if and only if there are at least two states which are
$E-$distinguishable. 
Note that hypothesis (ii) of the inheritance of superpositions theorem
ensures that the \me{}d \ob\ induced by the \mt\ scheme presented there is
a non-trivial \ob.

We now have all concepts in hand for stating the generalization of the
insolubility theorem for unsharp observables.

\proclaim{Insolubility Theorem for Unsharp Observables} Let $E$ be a
non-trivial \pov. There is no $E-$discrimination (and hence no $E-$\mt)
$\langle\hi_a,Z,T_a,U\rangle$ such that for all initial \s\ states $T$ the
\sa\ states $U\big(T\otimes T_a\big)U^{-1}$ are mixtures of eigenstates of
the projections $I\otimes Z(X_i)$ for a given reading scale $(X_i)$.
 \endproclaim

\demo{Proof}  
 First we show that for a non-trivial $E$ there exist pairs of orthogonal
vectors which are $E-$distinguishable. Indeed assume that for
any pair of orthogonal unit vectors $\{\fii_1,\fii_2\}$ one has
$p_{\fii_1}^E=p_{\fii_2}^E$. Then also the pairs 
$\big\{\frac 1{\sqrt 2}(\fii_1+\fii_2),\frac 1{\sqrt
2}(\fii_1-\fii_2)\big\}$ and $\big\{\frac 1{\sqrt 2}(\fii_1+i\fii_2),
\frac 1{\sqrt 2}(\fii_1-i\fii_2)\big\}$ are $E-$indistinguishable. From
this one infers that $\ip{\fii_1}{E(X)\fii_2}=0$. Since $\fii_1,\fii_2$ can
be arbitrary members of any orthonormal basis,  it follows that $E(X)$ is
diagonal in each such basis and hence $E(X)=\lambda(X)I$ for all
$X\in\Sigma$. Thus $E$ would be trivial, which was excluded.$^{12}$
 Suppose $\fii_1,\fii_2$ are two orthogonal unit vectors and
$E-$distinguishable. Suppose  $U\big(P_{\fii_1}\otimes T_a\big)U^{-1}$ and
$U\big(P_{\fii_2}\otimes T_a\big)U^{-1}$ are $I\otimes Z-$distinguishable
and are both expressible as mixtures of eigenstates of the projections  
$I\otimes Z(X_i)$.  Then the conditions of the inheritance of
superpositions theorem are satisfied. Therefore 
$U\big(P_\fii\otimes T_a\big)U^{-1}$ is not a mixture of eigenstates of
the $I\otimes Z(X_i)$, if $\fii$ is the superposition 
$c_1\fii_1+c_2\fii_2$ with nonzero
coefficients $c_1,c_2$. On the other hand, if vectors $\fii_1,\fii_2$ with the
assumed properties do not exist, then  the claim of the present theorem 
would also hold.
\enddemo

We remark as a by-product that this result encompasses all \pov{}s that may
be introduced for other reasons than dealing with inaccurate \mt{}s,
such as joint \ob{}s for noncommuting collections of \ob{}s.

\subheading{4. Conclusion}

\noindent
 We have extended earlier insolubility results in two ways: first, we have
taken the object observable to be an arbitrary \pv\ (sharp observable), and
second we have taken it to be an arbitrary \pov\ (unsharp observable when
not a \pv). A further step towards more realistic measurement situations
would be to consider measurement schemes where the pointer observable
$Z$ itself is an unsharp observable.
It is a largely open question whether an insolubility theorem can
be proved in such cases.$^{13}$

\subheading{Acknowledgements}

\noindent
 The research of one author (AS) was supported by the National Science
Foundation under grant $\#$PHY-93-21992. This work was carried out during 
PB's stay at Harvard University, funded by a Feodor-Lynen-Fellowship of
the Alexander von Humboldt-Foundation.
\vfill\eject

\subheading{Footnotes and References}

\item{\ 1.} 
E.~P.~Wigner, {\it Am.\ J.\ Phys.} {\bf 31}, 6 (1963).

\item{\ 2.}
B.~d'Espagnat, {\it Nuovo Cim.\ Suppl.} {\bf 4}, 828 (1966), and
{\it Conceptual Foundations of Quantum Mechanics}, Benjamin, Menlo Park,
1971, pp.\ 331-339.

\item{\ 3.}
A.~Fine, {\it Phys.~Rev.~D} {\bf 2}, 2783 (1970).

\item{\ 4.} 
A.~Shimony, {\it Phys.~Rev.~D} {\bf 9}, 2321 (1974).

\item{\ 5.}
{P.~Busch, P.~Lahti, P.~Mittelstaedt,} {\it The Quantum Theory of 
Measurement}, Springer-Verlag, Berlin, 1991. 2nd ed.\ in press, 1996.

\item{\ 6.}
H.~Brown, {\it Found.\ Phys.} {\bf 16}, 857 (1986).

\item{\ 7.}
A.~Fine, in {\it Kelvin's Baltimore Lectures and Modern Theoretical
Physics}, R.~Kargon and P.~Achinstein (eds.), MIT Press, Cambridge, MA,
1987, pp.\ 491-506.

\item{\ 8.}
{A.~Shimony}, {\it Search for a Naturalistic World View, Vol.\
II}, Cambridge University Press, Cambridge, UK, 1993, Section 2, pp.\ 41-47.

\item{\ 9.} H.~Stein, On a Theorem in the Quantum Theory of Measurement,
in {\it Experimental Metaphysics -- Quantum Mechanical Studies in Honor
of Abner Shimony}, D.\ Reidel, Dordrecht, to appear.

\item{\ 10.}
P.~Busch, M.~Grabowski, P.~Lahti, {\it Operational Quantum Physics},
Springer-Verlag, Berlin, 1995.

\item{11.}
In Ref.\ 10 the term {\it unsharp observable} is introduced in a somewhat more
restrictive sense, which can be ignored for  the present purpose.

\item{12.}
Note that for a Hilbert space of dimension greater than 2 there is a
simpler argument: for any two unit vectors which are not collinear there 
is at least one unit vector which is orthogonal to both. Then it is
obvious that $E-$indistinguishability for orthogonal pairs of vectors
implies $E-$indistinguishability for arbitrary pairs.

\item{13.} P.~Busch, Can `Unsharp Objectification' Solve the
Quantum Measurement Problem?, preprint, 1996.

\bye